\documentclass{PoS}

\title{The NICA/MPD Project at JINR (Dubna)}

\ShortTitle{The NICA/MPD Project at JINR (Dubna)}

\author{\speaker{Viacheslav Toneev }\\%
        JINR, Dubna\\
        E-mail: \email{toneev@theor.jinr.ru}}

\author{ for the NICA/MPD working group\\ }

\abstract{A new project of Nuclotron-based Ion Collider fAcility
(NICA) and Multi-Purpose Detector (MPD) is proposed at JINR
(Dubna). Status of this NICA/MPD project is outlined. Main system
parameters and some construction details are given.
          }

\FullConference{Critical Point and Onset of Deconfinement   -  4th
International
Workshop\\
                 July 9 - 13, 2007\\
                 Darmstadt, Germany}

\begin{document}

\section{Introduction}

The Joint Institute for Nuclear Research (JINR) in Dubna is an
international research organization established in accordance with
the intergovernmental agreement of 11 countries in 1956. At the
present time, eighteen countries are the JINR Member States and
five more countries have the associated member status. The JINR
basic facility for high energy physics research is represented by
the 6 AGeV Nuclotron which has replaced the old weak-focusing 10
GeV proton accelerator Synchrophasatron. The first relativistic
nuclear beams with the energy of 4.2 AGeV were obtained at the
Synchrophasotron in 1971. Since that time the study of
relativistic nuclear physics problems has been one of the main
directions of the JINR research program.

\begin{figure}[thb!]
\hspace*{30mm}\includegraphics[width=.6\textwidth]{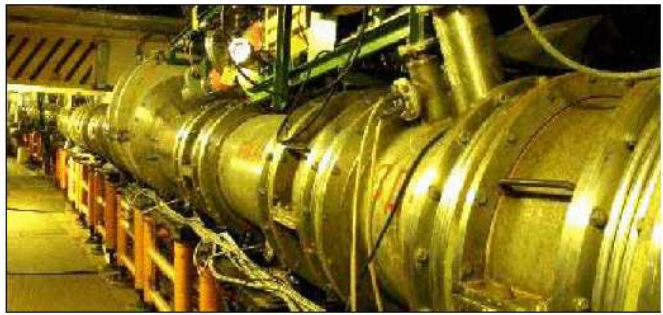}
\caption{JINR Nuclotron }
 \label{Nucl}
\end{figure}

The Nuclotron, 6 AGeV synchrotron, based on unique fast-cycling
supper-ferric magnets was designed and constructed at JINR
(1987-1992) and commissioned in March 1993. The annual running
time of 2000 hours was provided during the last years. Ion beams
up to iron ions and polarized deuterons  have been accelerated and
extracted from the accelerator. The view of the Nuclotron ring in
the tunnel is shown in Fig\ref{Nucl}.

The new flagship of the Joint Institute for Nuclear Research is
the NICA/MPD project to start experimental study of hot and dense
strongly interacting matter at the new JINR facility~\cite{SST06}
in the coming years. It may be reached by 1) Development of the
existing Nuclotron accelerator facility as a basis for generation
of intense beams over atomic mass range from protons to uranium
and light polarized ions; 2) Design and construction of heavy ion
collider (NICA) with maximum collision energy of $\sqrt s = 9$
AGeV and average luminosity $10^{27} \ cm^{-2}s^{-1}$, and 3)
Design and construction of a multi-purpose particle detector (MPD)
at intersecting rings. Realization of the project will lead to
unique conditions for the world community research activity. The
NICA energy region is of great interest because the highest
nuclear (baryonic) density under laboratory conditions can be
reached there. Generation of intense polarized light nuclear beams
aimed at investigation of polarization phenomena at the Nuclotron
is foreseen. Future JINR facility was  discussed at the Round
Table meeting "Searching for the mixed phase of strongly
interacting matter at the JINR Nuclotron"~\cite{RT1}.   A
conceptional project "Design and construction of the {\bf
N}uclotron-based {\bf I}on {\bf C}ollider f{\bf A}cilitiy ({\bf
NICA} and {\bf M}ulti-{\bf P}urpose {\bf D}etector ({\bf MPD}) was
first presented and discussed along with external experts at the
Round Table II in October 2006~\cite{RT2}.

The program of the Nuclotron upgrade in the NICA context is in
progress. The ions up to Au have been obtained  at test bench
based on the unique technology of highly charged state ion
sources. Modernization of the Nuclotron is one of the key points
in the NICA realization. We are planning the completion of the
first stage work by fall of 2009.

\section{NICA/MPD goals and physics problems}
\begin{figure}[thb!]
\hspace*{30mm}\includegraphics[width=90mm,clip]{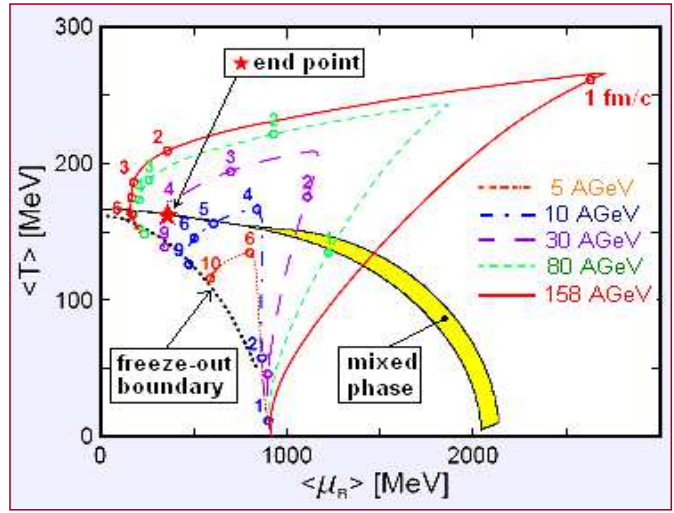}
\caption{Dynamical trajectories for central ($b=2$ fm) Au+Au
collisions in the $T-\mu_B$ plane for various bombarding energies
calculated within  the relativistic 3-fluid
  hydrodynamics with hadronic EoS~\cite{IRT05}. Numbers
  near the trajectories are the evolution time moments.  Phase
  boundary is estimated in a two-phase bag model taking into account
  the conservation of baryon and strangeness charges~\cite{KSTR06}.
  The critical end point calculated
  in the lattice QCD~\cite{Fodor01} is marked by the star.}
 \label{T-mu}
\end{figure}

Such an investigation is relevant for understanding the evolution
of the Early Universe after Big Bang, formation of neutron stars,
and physics of heavy ion collisions. The new JINR facility will
make it possible to study {\bf in-medium} properties of hadrons
and nuclear matter of the {\bf equation of state}, including a
search for possible signatures of deconfinement and/or chiral
symmetry restoration {\bf phase transitions}, and {\bf QCD
critical endpoint} in the region of $\bf \sqrt s_{NN}=3-9 \ GeV$
by means of careful {\bf scanning} in beam energy and centrality
of {\bf excitation functions}. As is seen from Fig.\ref{T-mu}, the
chosen energy interval indeed covers the most attractive domain of
the phase diagram.

\begin{figure}[t]
\hspace*{30mm}
\includegraphics[width=100mm,height=60mm,clip]{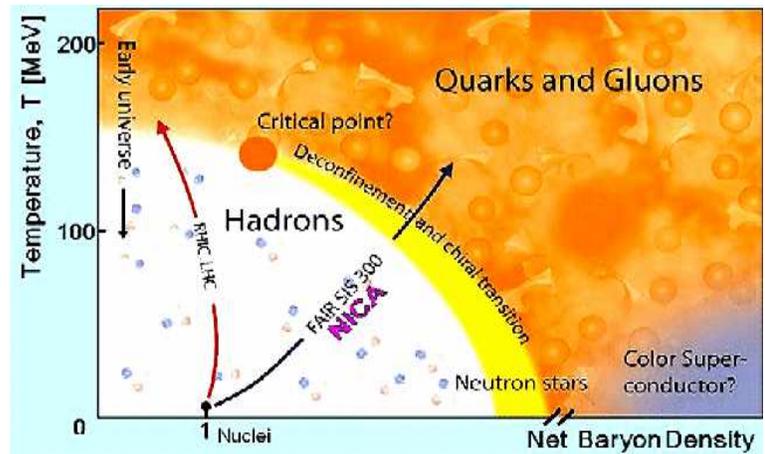}
\caption{Phase diagram (Artist's view).}
 \label{diag}
\end{figure}

The first stage measurements include:
\begin{itemize}
\item Multiplicity and global characteristics of identified
hadrons, including multi-strange particles;

\item Fluctuations in multiplicity and transverse momenta;

\item Directed and elliptic flows for various hadrons;

\item HBT and particle correlations.
\end{itemize}
Electromagnetic probes (photons and leptons) are supposed to be
added at the second stage of the project. The beam energy of the
NICA is very much lower than the energy region of the RHIC and LHC
but it sits right on the top of the region where the baryon
density at the freeze-out is expected to be the highest. In this
energy range the system occupies a maximal space-time volume  in
the mixed quark-hadron phase (the phase of coexistence of hadron
and quark-gluon matter similar to the water-vapor coexistence
phase). The net baryon density at LHC energies is predicted to be
lower. The energy region of NICA as well as of FAIR will allow
analyzing the highest baryonic density under laboratory conditions
(see the diagram in Fig.\ref{diag}). These conditions are expected
to be also at the FAIR facility in Darmstadt in 2015 when the
synchrotron SIS300 will come into operation.

\section{NICA general layout}
\begin{figure}[thb] \hspace*{30mm}
\includegraphics[width=100mm,height=60mm,clip]{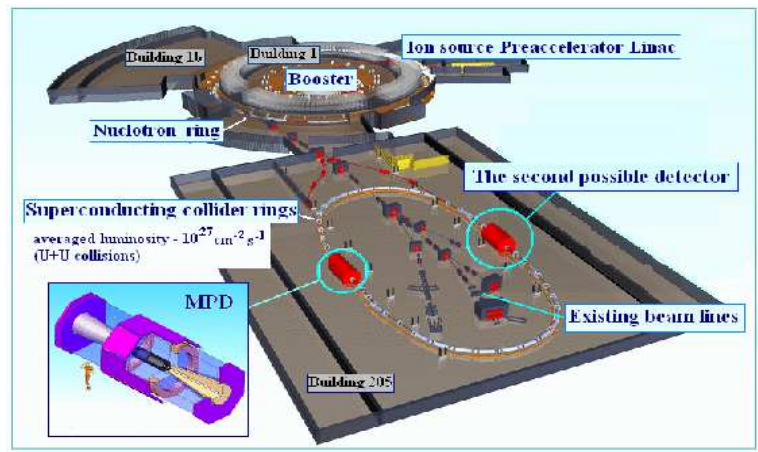}
\caption{General layout of the NICA/MPD }
 \label{gl}
\end{figure}

Construction of the new facility is based on the existing
buildings and infrastructure of the Synchrophasotron/Nuclotron of
the JINR Veksler-Baldin Laboratory of High Energies. The
accelerator chain includes (see Fig.\ref{gl}): Heavy-ion source -
Linac - Booster ring - Nuclotron - Superconducting collider rings.
The peak design kinetic energy of $^{238}U$ ions in the collider
is 3.5 AGeV. Beam cooling and bunching systems are foreseen. The
collider magnetic system is fitted to the existing buildings. The
project design presumes the use of some fixed target experiments.
A polarized deuteron beam from the Nuclotron will be available
too. The collision mode is under discussion. Several schemes of
the NICA complex have been considered since 2006. This one under
careful analysis at the present time is presented below.
\begin{figure}[t]
\hspace*{15mm}
\includegraphics[width=120mm,height=80mm,clip]{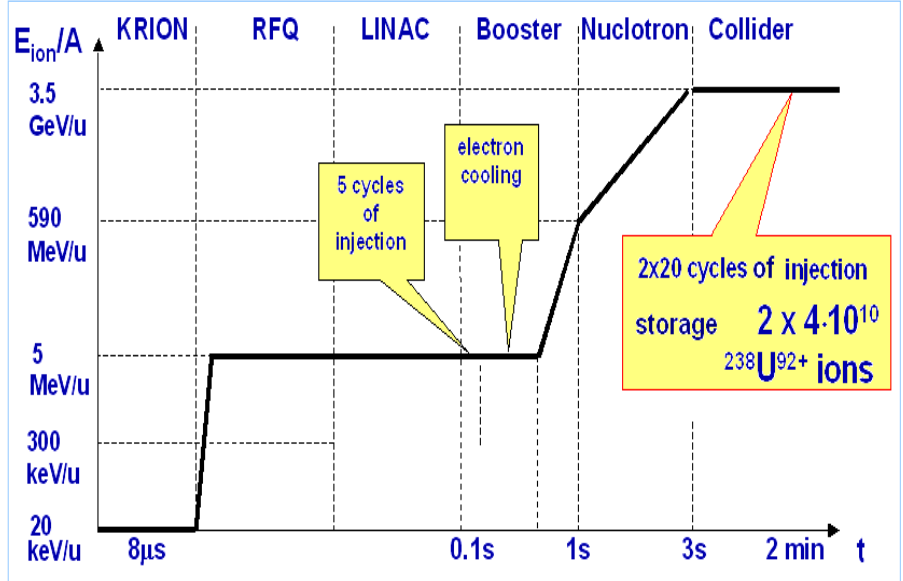}
\caption{The NICA time scheme}
 \label{tsc}
\end{figure}

\begin{table}[h]\hspace*{30mm}
\begin{tabular}{|l|rr|} \hline
Ring circumference, $m$&251.2&\\ 
Ion kinetic energy, $E \ [GeV/u]$, min/max&1 \ / \ 3.5& \\
Particle number per bunch, $N_{ion/bunch}$&2.0$\cdot 10^9$&\\
Bunch number, $n_{bunch}$& 20&\\
Horizontal emittance, $\varepsilon$ [$\pi \ mm \ mrad$]& 0.7&\\
Momentum spread, $\Delta p /p$& 0.001 &\\
IBS life time. [$sec$] & $\geq$ 100 &\\
Beta function at interaction points, $\beta^*$ [$m$] & 0.5&\\
RF voltage, $U_{RF}$ [$kV$] & 200 &\\
Laslett tune shift, $\Delta Q$& 0.0044 &\\
Beam-beam parameter & 0.009 &\\
Luminosity, $L$ [$cm^{-2}s^{-1}$] peak/average&2 \ / \ (1-1.5)$\cdot 10^{27}$ &\\
 \hline
\end{tabular}
\caption{The main NICA parameters } \label{tabl1}
\end{table}

Operation scenario is illustrated by the diagram in Fig.\ref{tsc}.
During the first 8 $\mu sec$ the KRION source produces the ion
beam with energy 20 $KeV/u$ which is quickly accelerated by RF
Quadruples till 5 $MeV/u$ and then after LINAC is injected into
Booster. Here the beam is cooled down electronically and is
accelerated again reaching the energy of 590 $MeV/u$ by the time
of 1 $sec$ completing the pre-injection stage. Now the beam may be
accelerated in the Nuclotron to the maximal project energy 3.5
$GeV/u$ and 20 cycles storage in every Collider Ring for about 200
$sec$ to be compatible with the Intra Beam Scattering (IBS)
lifetime. These 2$\times$20 cycles of injection result in 2$\times
4\cdot 10^{10}$ ions. The expected parameters of the collider are
presented in Table \ref{tabl1}. The specified average luminosity
of $1\cdot 10^{27} cm^{-2}s^{-1}$ can be reached at the chosen
parameters.

\begin{figure}[t]
\hspace*{15mm}
\includegraphics[width=120mm,clip]{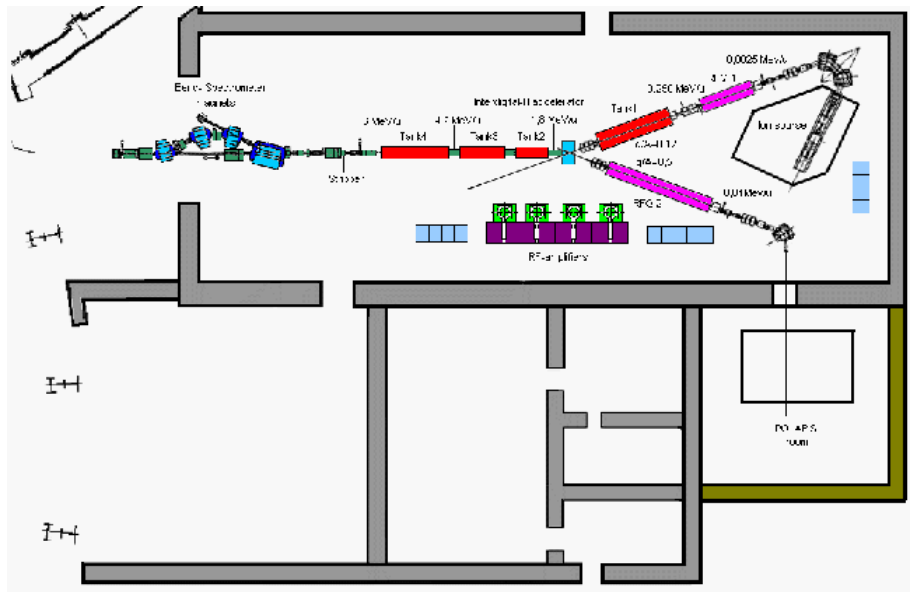}
\caption{Injector scheme}
 \label{linac}
\end{figure}

Let us consider the constituent elements of the NICA system in
more detail. The whole injection system is shown schematically in
Fig.\ref{linac}. Its heart element, the KRION ion source,  was
created at JINR~\cite{KRION} and, as is seen from Table
\ref{tabl2}, has a very significant advantage in the main
parameter, as compared to the ECR source: A number of ions per
unit time is higher by factor 5 in the KRION source. Note that for
the KRION all numbers in this table are measured experimentally.

\begin{table}[thb]\hspace*{30mm}
\begin{tabular}{|l|c|cc|} \hline
Ion source&KRION, $Au^{30+}$& ECR, $Pb^{27+}$&\\
\hline Peak ion current [$mA$] & 1.2 & 0.2 & \\
Pulse duration [$\mu sec$] & 8& 200& \\
Ions per pulse  & 2$\cdot 10^9$ & 1$\cdot 10^{10}$ &\\
Ions per $\mu sec$ &2.5 $\cdot 10^8$ & 5 $\cdot 10^7$ &\\
Norm. rms emittance & 0.15--0.3 & 0.15 -- 0.3 &\\
Repetition rate & 60 & 30 &\\
 \hline
\end{tabular}
\caption{Comparison of ion source parameters } \label{tabl2}
\end{table}

\begin{figure}[thb]
\hspace*{50mm}
\includegraphics[width=50mm,clip]{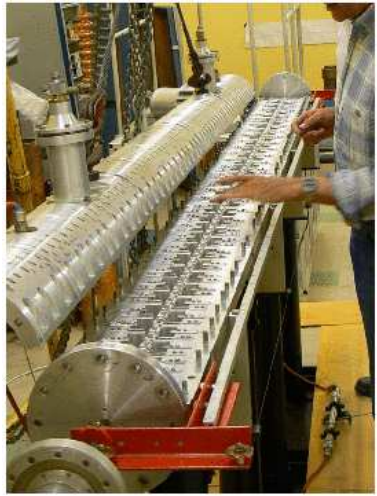}
\caption{Injector element at IHEP}
 \label{lin1}
\end{figure}

The Injector consists of Linac structures, RFQ and RF generator
(see Fig.\ref{linac}) and, in addition, Diagnostic, Control and
Water cooling systems  which will be manufactured at IHEP
(Serpukhov) and assembled at JINR. A similar system has been
constructed by IHEP for CERN (see Fig.\ref{lin1}). The cost of the
whole equipment is 10 $\$ M$.

\begin{figure}[thb]
\hspace*{20mm}
\includegraphics[height=45mm,clip]{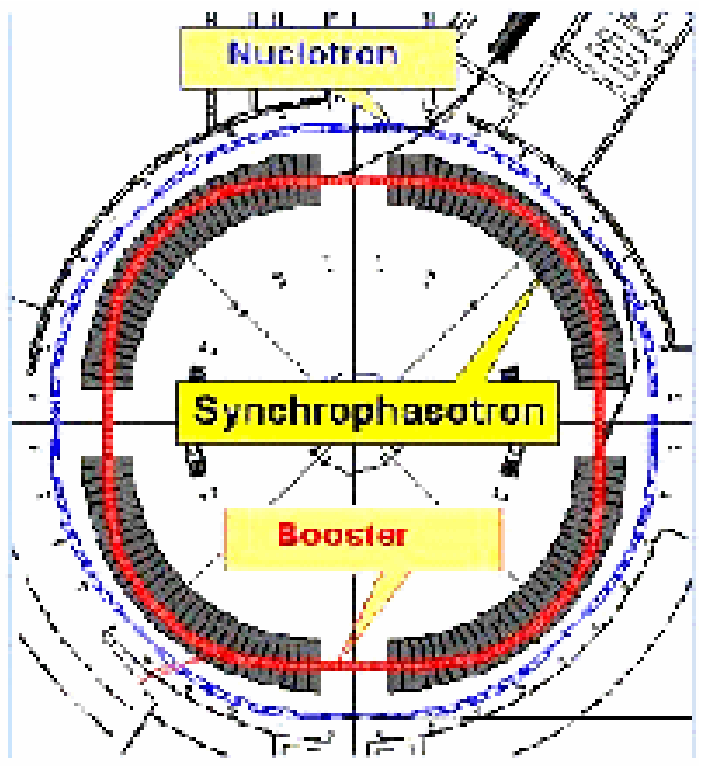}\hspace*{10mm}
\includegraphics[height=45mm,clip]{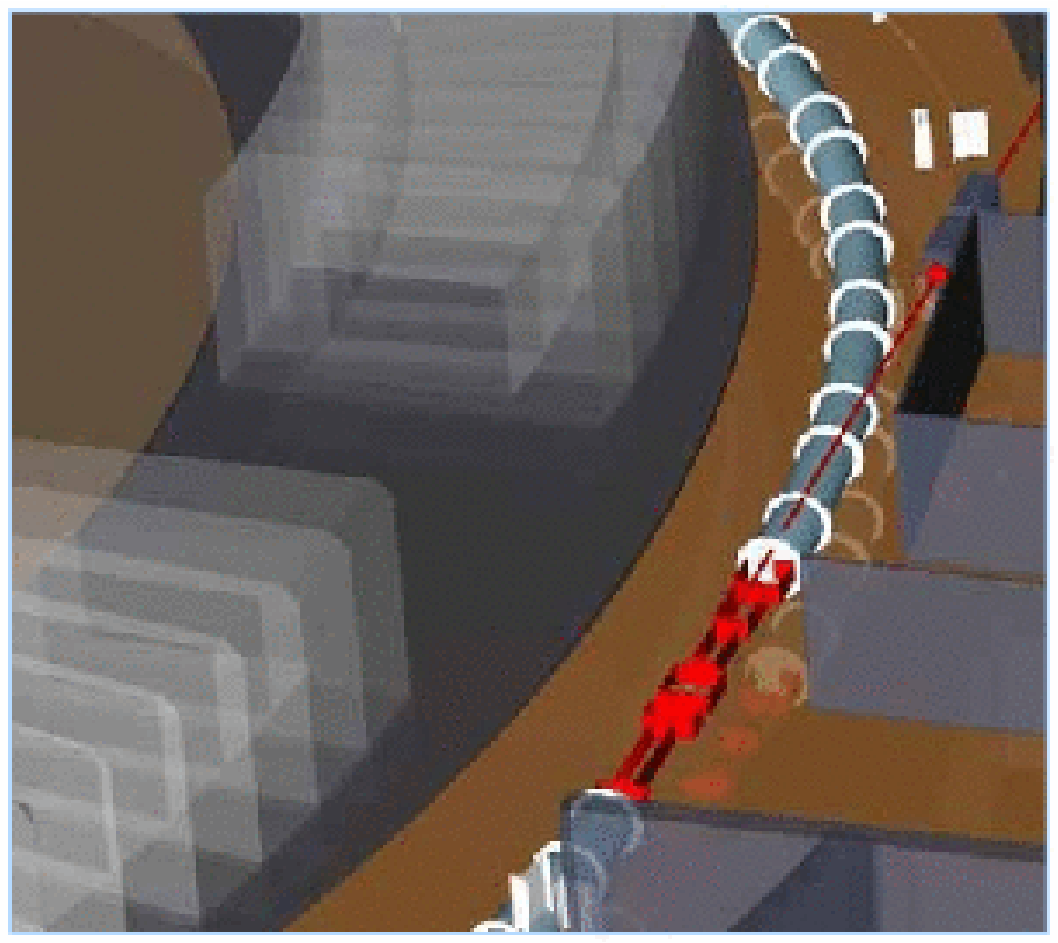}
\caption{Schematic (left) and pictorial (right) view of the
booster. }
 \label{b1}
\end{figure}

We consider the possibility to use a space inside the
Synchrophasotron magnet yoke window for installation of the
booster synchrotron, as is shown in Fig.\ref{b1}. The "warm"
Booster will have 70 dipole magnets of 1.37 T max magnet field. A
cost estimate of the booster is about 8 $\$ M $. It will be
manufactured by the Central Machinery Workshop of JINR.

The booster will fill two Collider Rings of about 250$m$
circumference. These rings have two intersecting points where
detectors will be installed.

 Note that the project has essential cost saving
factors: no new buildings, no additional power lines, and no extra
heat, water, cooling power are needed.

So, a NICA cost estimate (in $\$ M$): KRION+HV "platform" -- 0.25,
 Injector (IHEP design) -- 10, Booster -- 8, Collider Rings --
2$\times $10, {\it i.e.} totally  $\sim$ 40 $\$ M$.

\section{MPD for mixed-phase experiments}
\begin{figure}[thb]
\hspace*{5mm}
\includegraphics[width=140mm,clip]{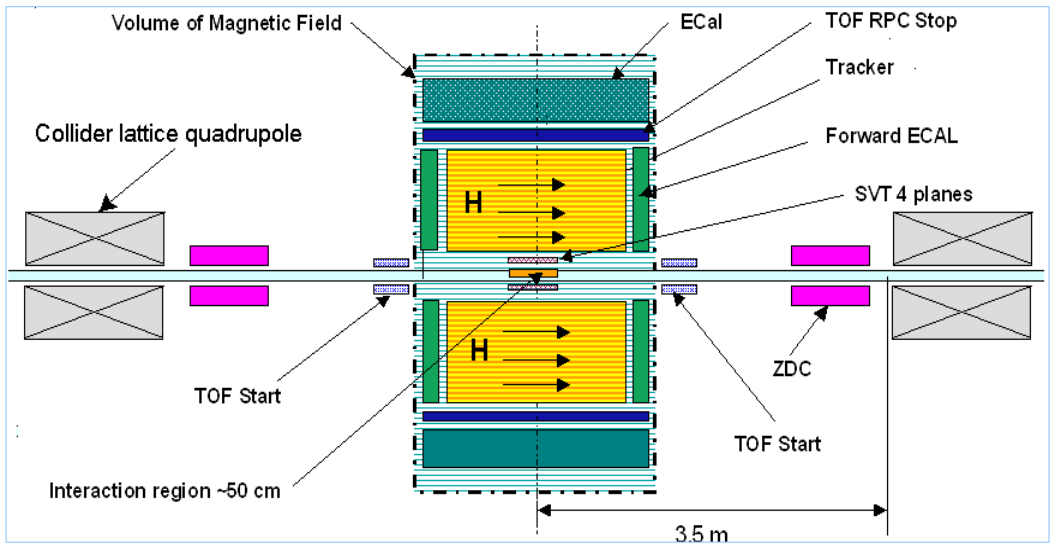}
\caption{General scheme of the MPD. }
 \label{mpd}
\end{figure}

The experimental set-up of proposed MPD has to detect the high
multiplicity events and perform particle identification. The MPD
scheme presented in Fig.\ref{mpd} indicates general elements of a
typical collider detector which is at the initial stage of
conceptional design. The tracking system, both the inner detector
based on silicon strips (Silicon Vertex Tracker) and the Tracker
provide the reconstruction of primary and secondary vertices and
precise measurement  of particle momenta. The Tracker includes  -
Central Tracker Detector (CTD) and Forward Tracking Systems (FTS).
Two alternative possible versions are considered as the CTD: the
TPC and Straw Tracker (ST). All tracking detectors are installed
in the magnetic field  of $\sim 0.5 \ T$ which is parallel to a
beam direction. For the particle identification Time of Flight
(TOF) System based on the RPC is proposed. This system allows
pion, kaon and proton identification in the momentum region $0.2-2
\ GeV/c$. The TPC option of the tracker could provide in addition
particle identification by measuring its ionization energy loss
($dE/dx$). For electron-positron and gamma detection an
Electromagnetic Calorimeter (ECal) is considered for the central
region of two Forward ECal's for high pseudo-rapidity. Two options
for these calorimeters are under consideration: Crystal or
Sandwich of plastic and lead multi-layers. Two near the beam sets
of scintillator counters are used as start devices for TOF
measurements and on-line vertex positioning. The two Zero Degree
Calorimeters (ZDC) provide the energy measurement of spectators
and determination of "centrality" in the ion-ion collision.

General requirements to the detecting MPD system are the
following: $|y|<2$ acceptance and $2\pi$ continuous azimuthal
coverage; high tracking efficiency; adequate track length for
tracking, momentum measurement and particle identification; two
track resolution providing a momentum difference resolution of a
few $MeV/c$ for HBT correlation studies; the fraction of
registered vertex pions $>70\%$.
\begin{figure}[thb]
\hspace*{10mm}
\includegraphics[width=60mm,clip]{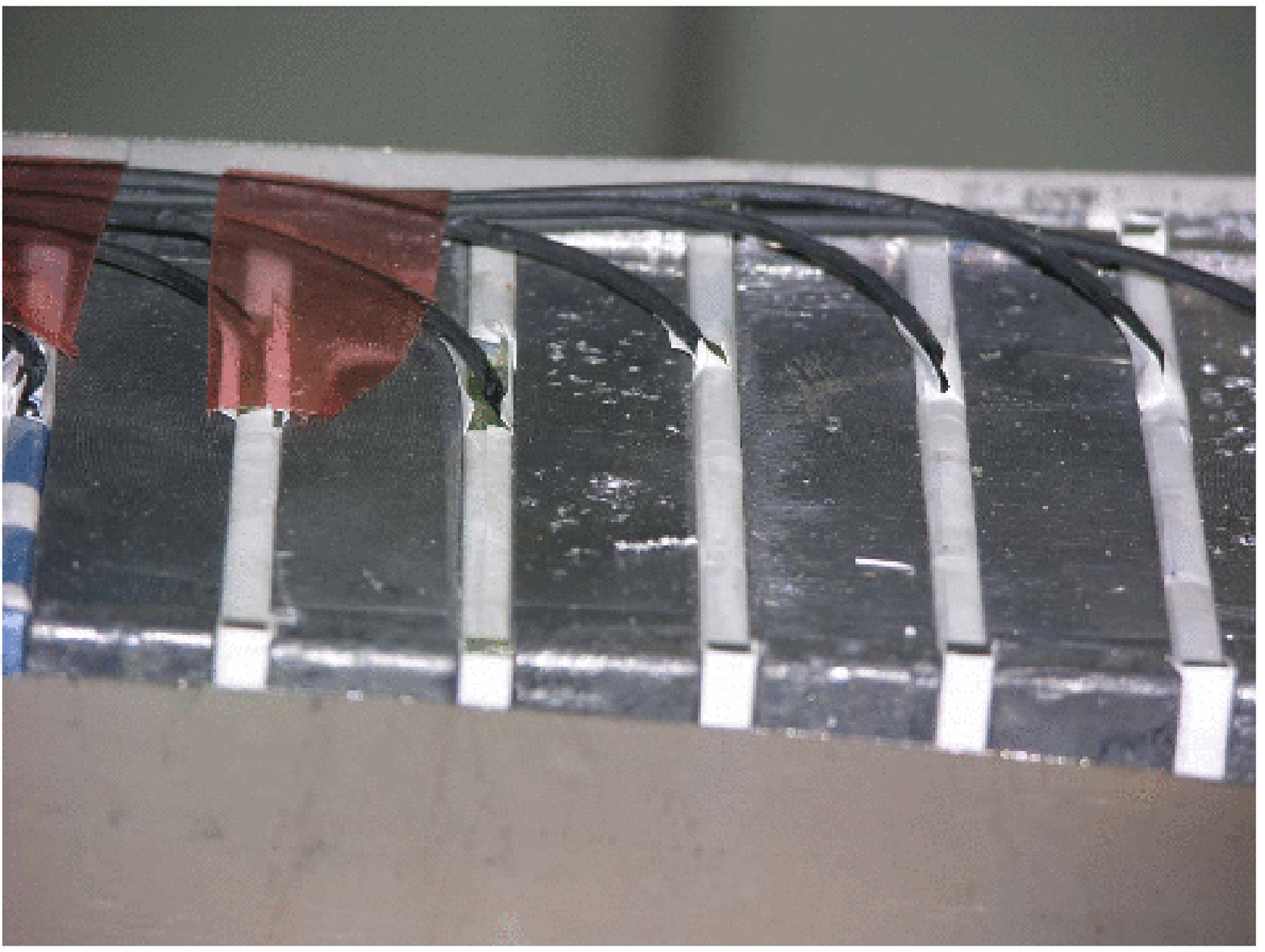}\hspace*{5mm}
\includegraphics[width=60mm,clip]{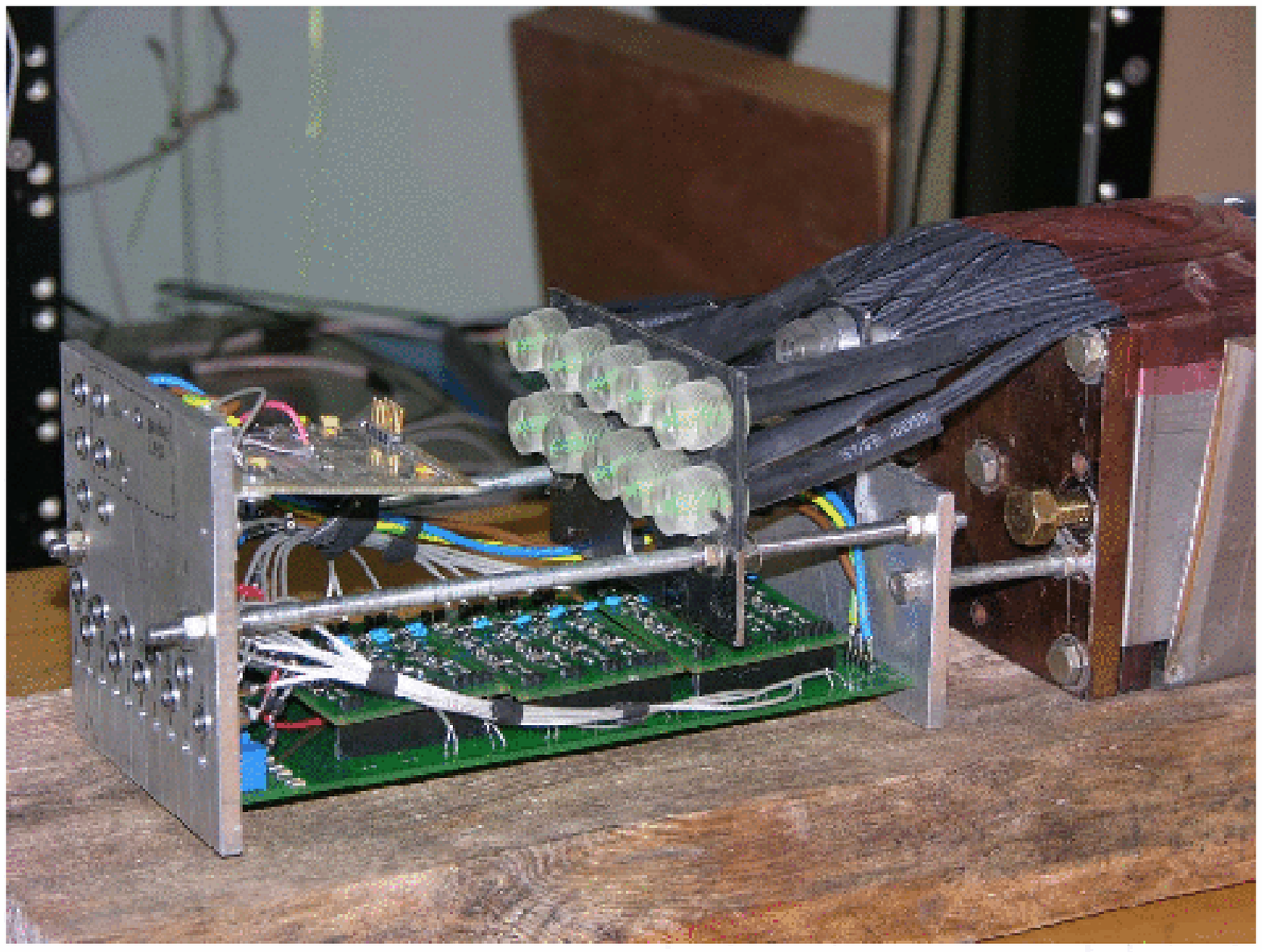}
\caption{ Elements of the ZDC at INR (Troitsk)}
 \label{zdc}
\end{figure}

Some basic parameters of the setup:
\begin{itemize} \item Interaction rate of $U+U$ events at
luminosity of $10^{27} \ cm^{-2}s^{-1}$ is 10 $kHz$; \item
Interaction rate of central events is 500 $s^{-1}$ (It is assumed
that DAC system is capable of reading-out and storing these
data);\item The accuracy of vertex positioning by means of the
silicon vertex detector is better than 0.2 $mm$; \item The TPC
produces 30 hits on a track and provides momentum measurement
accuracy of $\sim 1\%$ in the range of $p=0.2-2 \ GeV/c$; \item
The RPC time of flight system  has the time resolution of 100 $ps$
and provides $\pi - K$ separation  with probability $5\%$ for $p<2
\ GeV/c$. At low momenta $p<0.7 \ GeV/c$ all particle
identification is achieved by ionization measurement in TPC.
\end{itemize}

To select events, according to the centrality of a nucleus-nucleus
collision, Zero Degree Calorimeters will be constructed. They will
be made by INR (Troitsk) similarly to that manufactured for FAIR
 (see Fig.\ref{zdc}).

A crude estimate of the MPD cost (without Zero Degree Calorimeter
one) is about 25 $\$M$. In particular, it includes the cost (in
$\$ M$) of Silicon vertex detector -- 4.8, Time projection chamber
-- 5, TOF system -- 4, and EM calorimeter borrel -- 3.5.

\section{Summary and outlook}
The proposed stages of the NICA/MPD project realization are the
following:
\begin{itemize}
\item {\bf Stage 1: years 2007-2009}
\begin{itemize} \item Development of the Nuclotron facility
 \item Preparation of the Technical Design Report
 \item Start of prototyping of MPD and NICA elements
 \end{itemize}
 \item{\bf Stage 2: years 2008-2011}
\begin{itemize}\item Design and construction of the NICA and MPD
\end{itemize}
 \item{\bf Stage 3: years 2011-2012}
 \begin{itemize}\item Assembling
 \end{itemize}
 \item{\bf Stage 4: year 2013}
 \begin{itemize}\item Commissioning
 \end{itemize}
 \end{itemize}
 The total cost of the NICA/MPD project is about 70 $\$ M$.

Finally, the new facility at JINR in Dubna, the NICA/MPD, will
make it possible to study very important unsolved problems of
strongly interacting matter. The design and organization work has
been started. The Project Coordination Committee is established. A
special scientific department for preparation of the project is
formed. The first task of the NICA/MPD, the Conceptional Design
Report, is planned to be completed by this fall. We suppose a
wide-world cooperation with many Laboratories at both the R$\&$D
and construction stages of work.

\acknowledgments
 This work was supported in part by RFBR Grants N
05-02-17695 and 06-02-0400 as well as by the special program of
the Ministry of Education and Science of the Russian Federation
(grant RNP.2.1.1.5409).

\end{document}